\documentclass[pdflatex,sn-basic]{sn-jnl}

\usepackage{graphicx}%
\usepackage{multirow}%
\usepackage{amsmath,amssymb,amsfonts}%
\usepackage{amsthm}%
\usepackage{mathrsfs}%
\usepackage[title]{appendix}%
\usepackage{xcolor}%
\usepackage{textcomp}%
\usepackage{manyfoot}%
\usepackage{booktabs}%
\usepackage{algorithm}%
\usepackage{algorithmicx}%
\usepackage{algpseudocode}%
\usepackage{listings}%
\theoremstyle{thmstyleone}%
\theoremstyle{thmstyletwo}%

\theoremstyle{thmstylethree}%

\begin{document}

\title[Article Title]{Variable Red Giants Exploration with MASTER Robotic Net: Main Algorithms and Study of Three Mira-type Targets}


\author*[1]{\fnm{Alexander N.} \sur{Tarasenkov}}\email{tarasenkov@inasan.ru}

\author[1]{\fnm{Vladimir M.} \sur{Lipunov}}

\author[1]{\fnm{Artem S.} \sur{Kuznetsov}}

\author[1]{\fnm{Gleb A.} \sur{Antipov}}

\author[1]{\fnm{Pavel V.} \sur{Balanutsa}}

\author[1]{\fnm{Nataly V.} \sur{Tyurina}}

\author[1]{\fnm{Yakov Yu.} \sur{Kechin}}

\affil[1]{\orgname{Lomonosov Moscow State University}, \orgaddress{\street{GSP-1, Leninskie Gory}, \city{Moscow}, \postcode{119991}, \country{Russia}}}


\abstract{We are discussing the capabilities of MASTER network of robotic telescopes to study variable red giant stars. We report the discovery and present archival light curves for three optical transients, connected with long period variable red giants, based on MASTER Robotic Net all-sky survey observations and archival MASTER light curves at Lomonosov database storage. We demonstrate transient detection frames and light curves for three Mira-type variables: MASTER OT $J083717.54-573411.1$, MASTER OT $J190436.33+192828.7$ and MASTER OT $J07010810-6818548$.  Spectroscopically validated carbon star MASTER OT $J07010810-6818548$ shows mean magnitude growth but constant pulsation period which may indicate the presence of a long secondary period. As the result of MASTER wide-field images analysis at Lomonosov supercomputer data storage we present the 8-years historical light curves for all three objects with period calculations for all three Mira-type variables. We estimated fundamental astrophysical parameters for all three Mira-type variables: luminosities, radii, and zero age main sequence (ZAMS) masses.}

\keywords{late-type stars, variable stars, carbon stars, photometry, optical transients}



\maketitle

\section{Introduction}\label{sect:intro}

With the advent of wide-field survey telescopes, capable of obtaining light curves lasting tens of years with high accuracy for millions of stars, new opportunities have opened up for studying long-period variable red giant stars. Variable red giants include, for example, Mira variables and symbiotic stars (including the well-known recurrent nova T CrB)~\citep{TCRB, Masl, TCrb16, 2024MUPB...79..385T}. Not all mechanisms of these stars variability are sufficiently studied, and the effects detected during observations may be quite unusual and demonstrate non-typical characteristics for other types of stars. The variability amplitudes of these stars can be quite large and easily detected by wide-field survey robotic telescopes~\citep{OGLE2009}.

Mira type variables are Asymptotic Giant Branch (AGB) stars, which demonstrate long period and high amplitude pulsations~\citep{Mira_rev}. The variability of the prototype of Mira-type variables, the star Omicron Ceti (Mira), was discovered by Fabricius in 1596, but it is believed that some historical observations of “guest stars” were associated with the Mira variables~\citep{Hoffleit}. For the first time, the period of change in the brightness of Mira was estimated by Johannes Holwarda in 1638. ~\citep{Mirarad} was the first to estimate the radius of Mira Ceti. Spectral observations obtained at the beginning of the 20th century showed that Mira variables are giants of spectral class M with hydrogen emission lines and molecular absorption bands~\citep{Mira_spec}.

Currently, red giants with long pulsation periods and large variability amplitudes are classified as Mira variables. GCVS~\citep{GCVS} sets the following criteria to classify pulsating red giants as Mira variables: pulsation period longer than 100 days, and amplitude greater than 2.5 magnitudes in $V$ band. Infrared amplitudes are significantly smaller, than amplitudes in visual light. Similar objects with smaller amplitudes or periods are classified as long-period variables or semiregular variables. However, it is worth noting, that there is no clear physical boundary between Mira variables and semiregular variables, so for classification, not only formal criteria are taken into account, but also the observed behavior of the star. Mira variables are stars of relatively small masses ($M_{ZAMS}=1-3M_{\odot}$,~\cite{Fad23}) at a late stage of evolution after the helium flash, and their mass at the current stage of evolution is close to the solar mass~\citep{2004MNRAS.355..444I}. Initial mass ($M_{ZAMS}$) of the star is a key parameter of its evolution, including behavior at the AGB stage~\citep{Fad17}. Mira variables formed from solar-mass stars typically pulsate with periods of 200-250 days, while stars whose progenitors were more massive show significantly longer pulsation periods (400-600 days) and larger amplitudes~\citep{2012JAVSO..40..516W,2009ApJ...707..632L}. These stars belong to population II and are distributed in the galactic halo and bulge~\citep{Mira_kinem, Mira_kinem2}. The characteristic temperature of the Mira variables is about 3000K, and it changes greatly with the pulsation phase~\citep{2024Galax..12...81P}. Therefore, the color indices of these stars are red: $B-V$ is about $1.3-1.5$, and $J-K$ is more, than 2~\citep{2019BlgAJ..31..110Z, 2024arXiv240911266A}. At the moment of maximum brightness, the temperature is reaches maximum and the radius is minimal, and at minimum brightness, is the opposite situation~\citep{R_mir}. The processes in these stars are quite complex, include mass loss, dust particle formation, pulsations are possible both in the fundamental mode and in the overtone modes, but most of Mira variables are fundamental mode pulsators~\citep{Mira-mod, mir_mod2, 2024Galax..12...81P}. Therefore, the light curves of the Miras can have both a close to sinusoidal shape and quite unusual details~\citep{M-lc} and may experience significant long-term changes in the pulsation period~\citep{Mperiod1, Mperiod2}. Nevertheless, these variables demonstrate a good period-luminosity relationship, which can be used for estimation of astrophysical parameters~\citep{Periodlum}. Mira variables can be both O-rich and C-rich~\citep{BaylisAguirre2024} , which is clearly visible from the presence of characteristic molecular bands in their spectra ($VO$ and $TiO$ for O-rich and $C_2$ for C-rich)~\citep{O_rich}. Some atomic lines, for example hydrogen emission, strongly depend on the pulsation phase, disappearing in a minimum of brightness~\citep{2022AstBu..77..446P}. Mira variables can be part of symbiotic systems~\citep{simbio}, so their study is important for understanding the later stages of stellar evolution.

Here we present an analysis of the MASTER historical light curve for three such variable red giants, detected by MASTER Robotic Net~\citep{TNS_68} auto-detection system~\citep{Lip2019} during regular survey. MASTER OT $J07010810-6818548$ (J0701 hereafter) is the Mira-type variable, which shows a growth of the mean magnitude but a constant pulsation period. It is a spectroscopically validated carbon star~\citep{carbon1} that possibly can show outbursts like T CrB in the future. MASTER OT $J083717.54-573411.1$ and MASTER OT $J190436.33+192828.7$ (J0837 and J1904 hereafter correspondingly) are Mira-type variables. Main parameters of all three studied objects are presented in Table~\ref{param}.

\begin{table}[h]
\caption{Key parameters of objects under study.}\label{param}%
\begin{tabular}{@{}llllll@{}}
\toprule
Parameters & J0837 & J1904 & J0701 & Source \\
\midrule
R.A J2000.0  & 08:37:17.56 & 19:04:36.33 & 07:01:08.08 & \cite{Gaia3}\\
  Dec. J2000.0  & -57:34:10.6 & +19:28:29.0 & -68:18:54.8 & \cite{Gaia3}\\
  Var. type & M & M & M: & This work\\
  $log_{10}\frac{L}{L_{\odot}}$ estimate & $4.11\pm0.13$ & $4.14\pm0.13$ & $3.55\pm0.12$ & This work\\
  $\frac{M_{ZAMS}}{M_{\odot}}$ estimate & $3.0\pm0.1$ & $3.1\pm0.1$ & $1.2\pm0.1$ & This work\\
  $log_{10}\frac{R}{R_{\odot}}$ estimate & $2.68\pm0.04$ & $2.69\pm0.05$ & $2.33\pm0.04$ & This work\\
  Var. amplitude, mag & 3.2 & 4.1 & 2.7 & This work\\
  Var. period, d & $513.7\pm11.8$ & $501.6\pm7.5$ & $297.0\pm4.3$ & This work\\
  Pulsation mode & fundamental & fundamental & fundamental & This work\\
  Mean G mag  & 13.99 & 15.92 & 17.13 & \cite{Gaia3}\\
  Mean BP-RP color index  & 5.70 & 5.85 & 2.95 & \cite{Gaia3}\\
  Observations time span & 2015-2024 & 2012-2024 & 2015-2024 & This work\\
\botrule
\end{tabular}
\end{table}

\section{MASTER global network and main principles of its operation}\label{sect:master}

MASTER (Mobile Astronomical System of TElescope-Robots) global robotic net is a network of fully robotic wide-field twin telescopes at fast mount located around the globe, capable of observing of any point in the celestial sphere in one photometrical system at identical main equipment~\citep{Lipunov2010, Lip2019, Lip2022}. The main goal of MASTER robotic net is automatic detection of different optical transients~\citep{2024ARep...68.1364L}, including ones, connected with variable stars: eclipsing systems~\citep{Tyc}, microquasars~\citep{NewA404}, cataclysmic variables of different types~\citep{RedNova,WZSge,polar}, exoplanet search~\citep{MasterexoARep,Masterexo1, Masterexo2} and other photometrically variable objects. Key technical parameters of MASTER telescopes are presented in table~\ref{param_T}. MASTER network consists of 9 nodes, which are described in table~\ref{Mobs}. The location of telescopes around the globe allows observing objects without declination restrictions, and track interesting objects continuously.

\begin{table}[h]
\caption{Key parameters of MASTER telescopes.}\label{param_T}%
\begin{tabular}{@{}lll@{}}
\toprule
Parameter & Value & Comments \\
\midrule
  Telescope optical system  & Hamilton & \\
  Telescope aperture  & 400 mm & \\
  Telescope focal length  & 1000 mm & F/2.5 focal ratio\\
  Telescope FoV & 4 square degrees & 2{\textdegree}x2\textdegree\\
  Pointing speed & 30{\textdegree} per second &\\
  Tubes per mount & 2 & Both converged and separated tube modes are supported\\
  Detector & Apogee Alta U16M & \\
  Photometric system & BVRI + 2 polarization filter & 2 photometric + 1 polarization filter on each tube\\
  Polarisation filters & perpendicular oriented & for linear polarization~\citep{Pol2014, PolGRB} \\
  Limiting magnitude & 19-20$^m$ & Unfiltered band\\
  Survey speed & 480 square degrees per night & 2 frames per night for each field\\
\botrule
\end{tabular}
\end{table}

\begin{table}[h]
\caption{MASTER observatories.}\label{Mobs}%
\begin{tabular}{@{}ll@{}}
\toprule
Observatory name & Location \\
\midrule
  MASTER-Amur & Observatory of Blagoveshchensk Pedagogical University, Russia \\
  MASTER-Tunka & Tunka station of Irkutsk State University, Russia \\
  MASTER-Ural & Kourovka observatory, Russia \\
  MASTER-Kislovodsk & Kislovodsk Solar Station of Pulkovo observatory, Russia \\
  MASTER-Tavrida & Crimean astrophysical observatory \\
  MASTER-SAAO & SAAO observatory, South Africa \\
  MASTER-IAC & Institute of Astrophysics of the Canary Islands, Spain \\
  MASTER-OAFA & Felix Aguilar Observatory, Argentina \\
  MASTER-OAGH & Guillermo Haro Observatory, Mexico \\
\botrule
\end{tabular}
\end{table}

\subsection{MASTER observation algorithms}\label{Algor_M}

MASTER telescopes conduct observations in three modes: alert observations~\citep{2003GCN, Lipunov2003b,Bart1}, inspection, and regular sky survey~\citep{LipBook, Lip2024}. For alert observations two modes are usually used: alert (in the first hour after the events alerted by Swift, Fermi, GECAM, MAXI, Integral, Einstein Probe, LIGO/Virgo, IceCube, ANTARES/KM3Net) and inspection (if more than an hour has passed since the trigger).Observational strategy (duration, number of exposures) for every trigger is determined taking into account time from trigger, error box size, other active alerts and weather conditions. For events with error-box size less than a degree (Swift, Fermi-LAT, Integral missions), observations are carried out to a height of 0 degrees above the horizon in alert mode (when notice time is less than 30 minutes). The technical capabilities of the telescope mount allow observations at zenith distances of approximately 90 degrees and even slightly greater, which is important for detecting the optical emission of GRBs at low altitudes, especially in mountain observatories, where the observed horizon is lower than the mathematical one. For cases of large fields (Fermi-GBM, LIGO/Virgo, IceCube), observations are carried out in a minimum of two frames with a gap of 15 minutes in time between them. This is necessary to separate optical transients from moving objects and CCD matrix defects. If there are no active triggers, MASTER conducts a regular sky survey  according to the same scheme - at least 2 frames per night with limit control and offset between frame centers to automatically exclude hot pixels and cosmic particles from the list of transient candidates. The survey typically uses unfiltered (W-band) observations, exposures of 60-120 s with a limiting magnitude of 20$^m$ on moonless nights~\citep{Gorb2013}.

\subsection{Data reduction and transient detection system}\label{Data_Red}

MASTER network has its own online reduction and auto-detection system, that includes primary processing (BIAS, DARK, Flat Field correction), identification of all optical sources at every image, calculation of parameters for moving ones between neighboring images. In regular survey we image every field as minimum 2 times per night  for stationary objects search mode or 3 times per night for new asteroids and comets detection. To distinguish a candidate, it must be present in at least two frames obtained in one night, and absent or significantly different in brightness in the archival reference frame. Every candidate is checked in all main catalogs at VIZIER~\citep{vizier2000} database, TNS (transient name service) database and Minor Planet Center (MPC) to exclude known variable stars and Solar System objects. MASTER autodetection system can find candidates up to 20$^m$~\citep{Orphan_ARep, Gorb2013}.

Standard online MASTER reduction includes photometry, astrometry, and optical transients detection~\citep{Lipunov2010, Kornilov, Lip2023, Lip2019, Lip2022} after primordial automatic analysis. The main requirement for this reduction is the short time on full programs complex. It must be not more than 30-60 seconds for dozens of thousands stars at every image.
 
After detection each transient candidate is checked in the MASTER archive for light curve analysis (if any), in the VIZIER~\citep{vizier2000} database, The American Association of Variable Star Observers (AAVSO) variable star databases, catalogs of X-ray, UV, IR and radio sources, as well as in open archives of review projects such as ATLAS~\citep{atlas}, CRTS~\citep{crts}, ROTSE~\citep{rotse} and others. Any candidate can be checked manually using special tools~\citep{Lip2019}. The MASTER autodetection system is a universal tool and allows detecting optical transients of any type: optical afterglows of gamma-ray bursts, supernovae, flares of cataclysmic variables, AGN variability, antitransients (fading objects). Variable red giants such as Mira variables are easily detected by this system due to the large amplitude of variability.

\subsection{MASTER archive at Lomonosov database storage}\label{Data_Lom}

MASTER image archive is stored at Lomonosov data storage~\footnote{https://parallel.ru/cluster/lomonosov.html}~\citep{Voevodin}.
The complex is managed by the OpenNebula cloud computing platform~\citep{OpNeb1,OpNeb2}, and the data storage is managed by Ceph~\citep{Ceph} software; together, the above software allows flexible management of the complex and ensures scaling of all its components. For the global network of robotic telescopes, MASTER, a complex of virtualization of calculations and data, provides resources in the Infrastructure-as-a-Service (IaaS) format. 

MASTER archive at Lomonosov data storage and data virtualization complex contains all frames, obtained by MASTER network for all the time of its operation, including several thousands long series of observations, when one area in the sky was continuously observed for several hours~\citep{Masterexo2,Lipunov2010}. The main parameters are presented in table~\ref{dbase}.

\begin{table}[h]
\caption{Key parameters of MASTER database.}\label{dbase}%
\begin{tabular}{@{}lll@{}}
\toprule
Parameter & Value & Comments \\
\midrule
  Total frame number  & 6,000,000 & \\
  Total volume  & 100 Tb & \\
  Single frame volume  & 70 Mb & \\
  Frames obtained in & 2003-2025 & ~\cite{2003GCN}\\
\botrule
\end{tabular}
\end{table}

The centralized storage of this data allows researcher to simplify and significantly speed up access to the huge archive of images, formed during the entire operation of the robotic network for automatic processing and facilitate the manual search for the necessary information by the researcher. To solve these problems, a whole range of hardware and software is implemented, providing effective interaction of elements of the robotic network of telescopes, supporting continuous data exchange with each node of the network and providing operational access to the archive of scientific data for authorized users (dynamically integrated database). Data stored in the database can be obtained upon a motivated request at MASTER webpage~\footnote{https://observ.pereplet.ru/}.

\section{Variable stars detection by MASTER online reduction}\label{sect:var}

Variable stars are detected at every MASTER images~\citep{NewA404, Lip2022}. The red giants are quite common among them. The variability of such stars can be connected with interaction with companion star, pulsations, dust accumulation and discharge in the atmosphere, envelope dissipation  and other~\citep{RG1,RG2}. 

Thanks to the fact that during the operation of the MASTER network, a huge archive of images spanning more than 20 years of work has been formed, it has become possible to study the long-term photometric behavior of each discovered variable star. This is especially important for long-term variables, such as Mira-type variables, which require long observation series for detailed study. MASTER data allows not only the detection of such long-period objects, but also a detailed study of their astrophysical parameters. To demonstrate this, we are considering as an example three Mira-type optical sources, found by MASTER auto-detections system. We can obtain long light curves from Lomonosov data storage for all of them, that let us calculate these objects variability period. Below we present an analysis of these three Mira variables (J0837, J1904 and J0701) based on data from the MASTER archive.

For all the objects we discovered, we provide archival light curves, as well as a set of discovery frames. In the MASTER network survey, a two-pass scheme is usually used~\citep{Lip2022}. the discovery of a transient is considered confirmed if it is visible in frames from both passes. Therefore, for each object we present two discovery frames and an archive reference frame. All 3 frames are 6' by 6' cutouts generated by the transient confirmation system.

Since automatic photometry algorithms may not be accurate enough for detailed photometric analysis, we carried out processing of all frames in the specialized program AstroimageJ~\citep{AiJ}, developed specially for obtaining high precision light curves using the aperture photometry method. The size of the photometric aperture was calculated based on the FWHM of the target star. To improve the accuracy of the calculations, the characteristics of the telescope's CCD matrix (gain, readout noise), background inhomogeneity across the frame were taken into account. For frames obtained without a filter, G magnitude values from Gaia DR3~\citep{Gaia3} were taken as standard magnitudes. For $BVRI$ photometry, the standard values were also taken from~\cite{Gaia3}, from synthetic photometry dataset, based on Gaia low-resolution spectroscopy. During the processing, the photometric behavior of comparison stars was studied; variable stars were excluded from the analysis. For photometry of archive frames of all three studied objects we used field stars as comparison stars. The comparison stars were selected so that their Gaia-magnitude was close to the average value of the target star obtained by the MASTER automatic processing system. We filtered the obtained measurements by errors, discarding all measurements where the measurement errors exceeded $3\sigma$ from mean error-magnitude dependence for each filter installed on MASTER telescopes. The example of such error-magnitude dependencies is presented in~\cite{Gorb2013}.

For the initial period search we used the WinEfk software~\footnote{http://www.vgoranskij.net/software/WinEFengVers2015-09-08.zip}, which is based on methods, developed by~\citep{LKper} and~\citep{Dper}. The main algorithms of the WinEfk software are described in the work~\citep{Goran1}. Although this program does not provide the ability to determine period errors, due to its user-friendly interface and a large set of functions (calculation of periodograms using various methods, trend removal, construction of light curves, calculation of light elements, data cleaning, etc.) it finds wide application in variable stars investigation~\citep{2008AcA....58..279K,2015RAA....15.1005K,2024MNRAS.530.1328I}.

\subsection{Mira-type Optical Transient MASTER OT J083717.54-573411.1}\label{J0837}

\begin{figure*}[h]
\centering
\includegraphics[width=0.9\textwidth]{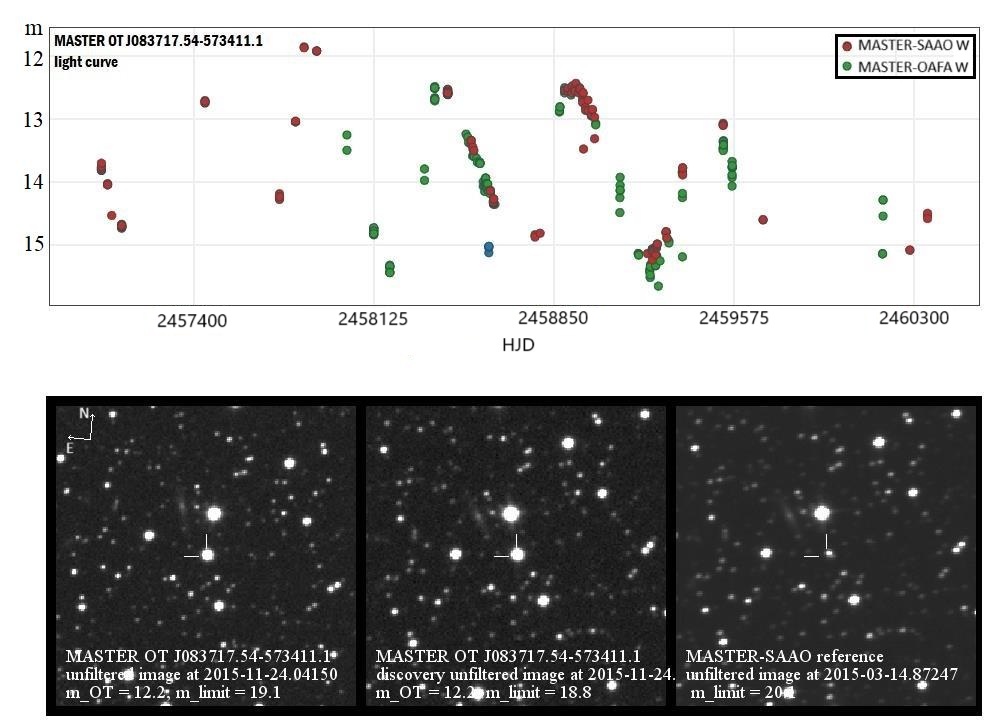}
\caption{An archive light curve of J0837 (upper panel, the red circles are from MASTER-SAAO, the green ones are from MASTER-OAFA in unfiltered band) and the OT discovery and confirming images with reference one - lower panels. Here and further, discovery and reference frames have a size of 6' by 6'.}\label{Fig1}
\end{figure*}

During regular survey variable star J0837 was found by MASTER-SAAO auto-detection system  at position (R.A., Dec. J2000) = 08h 37m 17.54s -57d 34m 11.1s on 2015-11-24.03137 UT with unfiltered magnitude 12.2 mag (the image limiting magnitude is 19$^m$.1) and was initially misclassified as a QSO flare~\citep{2015ATel.8335....1B}. MASTER-SAAO and MASTER-OAFA photometry for 9 years is presented in Fig.~\ref{Fig1}, demonstrating classical Mira-type variability. The  VIZIER analysis gives conclusions about this very red object ( $K$ = 5.3 mag; $J-K$ = 2.3 mag in 2MASS and $R$ = 14.9 - 18.0 mag in USNO-B1.0 catalogs). The object was identified with the infrared source IRAS 08360-5723~\citep{1994yCat.2125....0J} and a low-resolution IRAS spectrum was obtained for it~\citep{IRAS_LRS}. The object was classified as oxygen-rich AGB star~\citep{IRAS_SPclass}. According to Gaia, the object was classified as a long-period variable~\citep{2018A&A...618A..58M,2023A&A...674A..15L}.

Using WinEfk program we obtained pulsation with a period of 518.1 days and using Lomb-Scargle periodogram - $513.7\pm11.8$ days, which agrees well with the period determined from ATLAS photometric data~\citep{atlas}. J0837 demonstrates periodic light curve with amplitude of 4.0$^m$. The duration of the ascending branch is somewhat shorter than the descending one, which is typical for pulsations of Mira type variables. 

\subsection{Mira-type Optical Transient MASTER OT J190436.33+192828.7}\label{J1904}

 During regular survey J1904 was found by MASTER-Kislovodsk auto-detection system at (R.A., Dec. J2000) = 19h 04m 36.33s +19d 28m 28.7s on 2015-02-23.08345 UT with unfiltered magnitude  14.0 mag (the limit was 18.6 mag). The object was detected on 18 images received during the discovery night~\citep{2015ATel.7147....1S}. The infrared color index J-K is 2.98 mag (VIZIER database). The object was identified with the infrared source IRAS 19024+1923~\citep{1994yCat.2125....0J} and a low-resolution IRAS spectrum was obtained for it~\citep{IRAS_LRS}. The object was classified as oxygen-rich AGB star~\citep{IRAS_SPclass}. The object was classified as a long-period variable named OGLE~GD~LPV-24072 by OGLE survey~\citep{OGLE2022} and as a Mira-type variable by ZTF survey~\citep{ZTFVAR}. 
 
\begin{figure*}[h]
\centering
\includegraphics[width=0.9\textwidth]{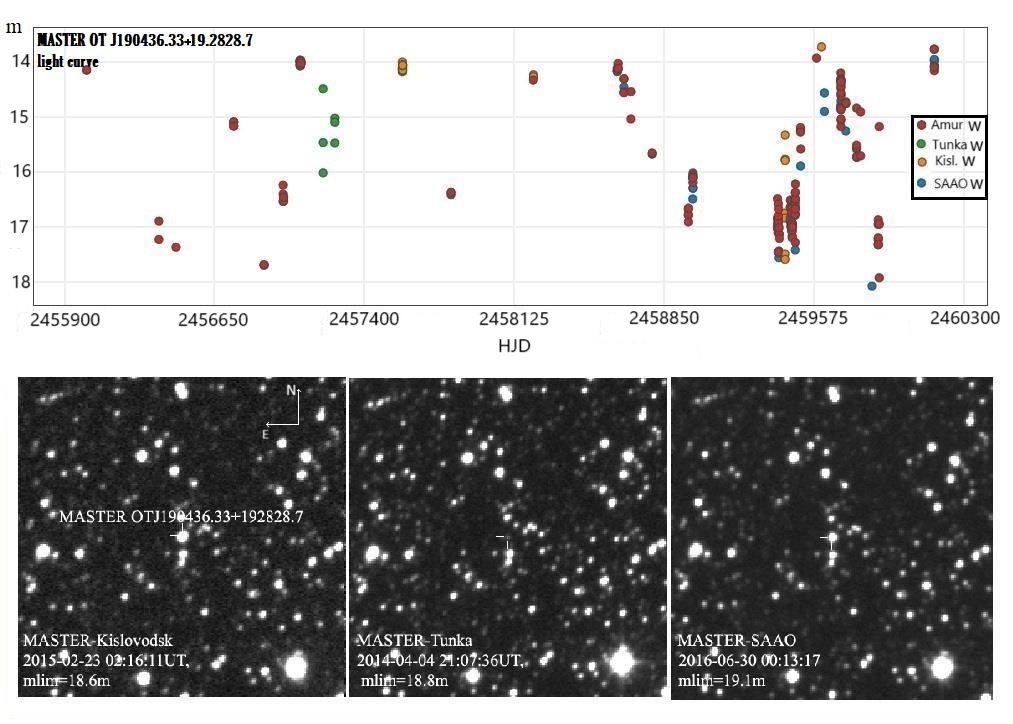}
\caption{An archive light curve of MASTER OT MASTER OT J190436.33+192828.7 (upper panel). The different colored circles are from MASTER-Amur (red), -Tunka (green), -Kislovodsk (orange), -SAAO (blue) in unfiltered band and the OT discovery and  confirming images with reference one (lower panels).}\label{Fig2}
\end{figure*}

 The MASTER historical light curve demonstrates the classical Mira-type one (See Fig.~\ref{Fig2}). Estimated period is 499.4 days (WinEfk) and $501.6\pm7.5$ days (Lomb-Scargle periodogram). The period determined from MASTER data is in good agreement with the periods determined from OGLE and ZTF data. J1904 demonstrates periodic light curve with amplitude of 4.1$^m$. The duration of the ascending branch is somewhat shorter than the descending one, which is typical for pulsations of Mira type variables.

\subsection{MASTER OT J07010810-6818548: Carbon Mira variable brightening}\label{J0701}

J0701 was detected in frames, obtained by MASTER-SAAO at 2024-03-11 UT 20:13:51.744~\citep{TNS_68} as 14.7 magnitude transient. These frames were part of GRB errorbox observation. The object was identified with a star 2MASS J07010810-6818548~\citep{2MASS}, which is a spectroscopically validated carbon star~\citep{carbon1, carbon2}. We conducted a detailed analysis of the behavior of this object using data from the MASTER network archive. It was found that this star exhibits pulsations with an amplitude of 2.7 magnitudes in unfiltered band and a period about 300 days. We concluded that this object belongs to the class of Mira-type variable stars. However, at the moment JD 2458800 the average brightness of the star began to increase while maintaining the same period and amplitude. By now it has changed from 17.0 to 15.5 magnitude in unfiltered band, and at its maximum the star has reached 15.0 magnitude. 

Analysis of the properties of J0701 showed that they are similar to some other Miras with long term trends~\citep{Mir1}. Mira variables can exhibit periodic and quasi-periodic trends in brightness, with characteristic times of tens of years, superimposed on the main pulsation curve. This behavior may be due to the presence of a companion star or planet slowly sinking into the rarefied atmosphere of the red giant~\citep{Rud1}.

We obtained $RI$ photometry for J0701 using MASTER-SAAO telescope. Mean $R$ magnitude in brightened state is $14.92\pm0.05$ and mean $I$ magnitude $13.57\pm0.05$. The $R-I$ color index of $1.35m$ corresponds to spectral class $M$, characteristic of carbon stars and Mira-type variables. Multicolor monitoring light curves show in some nights flickering-like effects with amplitude of 0.05 magnitudes and period less than a hour and brightness level variations on hours scale with amplitude of about 0.5 magnitudes in I. However, it cannot be said that this detection was reliable, since on the nights when this effect was recorded, the weather conditions were unstable (changes in FWHM, haze).

Analysis of TESS~\citep{TESS} photometry of J0701 shown, that this object does not demonstrate flickering or short-term photometric variability. Although flickering is typically detected at shorter wavelengths than TESS provides, TESS photometry allows this behavior to be detected in a number of cases~\citep{TFlick}. We cannot say with complete certainty that the absence of detected flickering in the TESS data means that J0701 does not exhibit flickering at all. Photometric monitoring during the minimum brightness in the U or B filter simultaneously with the longer-wave one, as for example in~\cite{2025NewA..12102452Z}, is necessary to reliably reveal or refute flickering.

\begin{figure*}[h]
\centering
\includegraphics[width=0.9\textwidth]{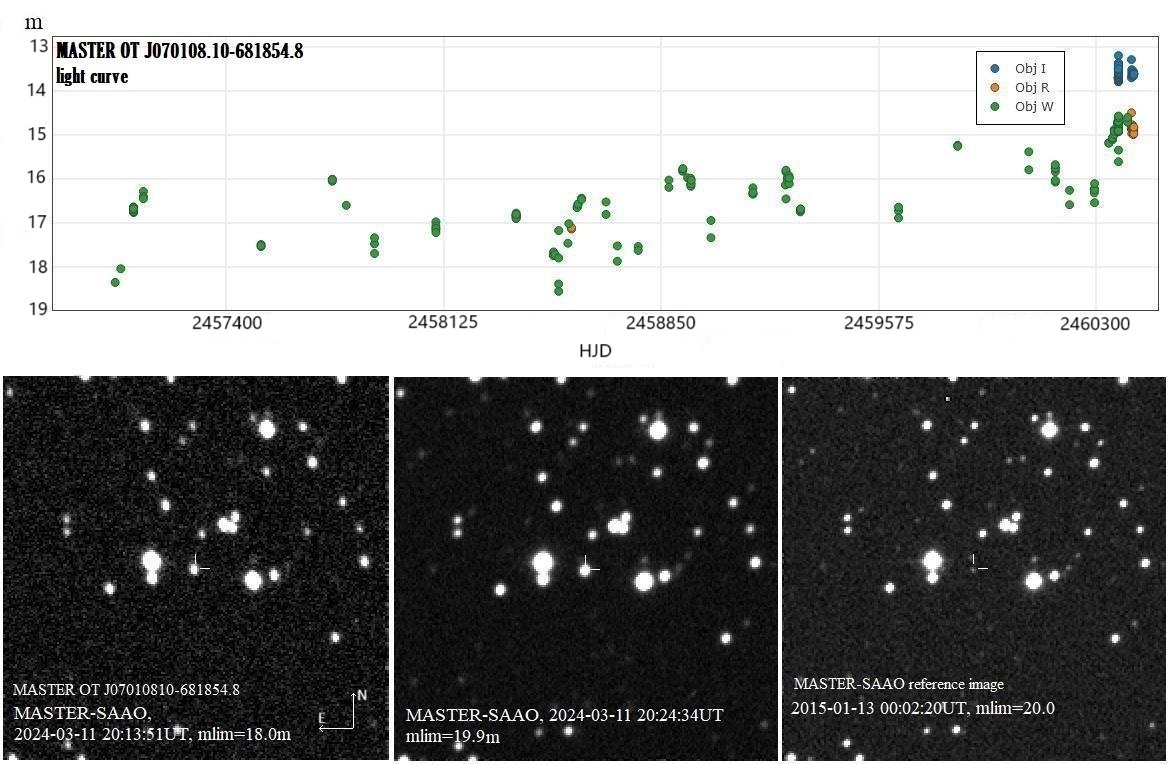}
\caption{Archive MASTER-SAAO light curve of MASTER OT J07010810-681854 in unfiltered band and discovery images in February and reference one in 2015.}\label{Fig3}
\end{figure*}

Archive light curve of J0701 in unfiltered band obtained by MASTER-SAAO telescope is represented in the Fig.~\ref{Fig3}. Using WinEfk software we obtained pulsation with a period of 293.5 days. Lomb-Scargle periodogram gives a period of $297.0\pm4.3$ days for detrended light curve. The pulsation period does not change significantly over the observation period, despite the increase in average magnitude. This makes it possible to use the derived period to determine the fundamental parameters of a star.

Interpretation of the observed trend in light curve of J0701 is a rather complex problem due to the complexity of the processes occurring in the atmospheres of red giants. Long term mean amplitude variations similar to those that we discovered were observed, for example, in S Aur~\citep{cadm24} and T Dra~\citep{tdra} light curves. This may be a manifestation of the so-called long secondary period variability - the presence of a brightness variation with a large amplitude, in addition to the main pulsation period. This variation has significantly longer period, than the pulsation one~\citep{lpv1, Iwanek25}. It is worth noting, that the long secondary period is not as common in Miras as in other long period variables, and there are no confirmed Miras with long secondary period. Also, Mira variables which are candidates to have long secondary period are mostly carbon~\citep{lpv2}. The physics of long secondary period is not completely understood, and two main hypotheses of its origin are presence of a planet or substellar companion~\citep{binmir} or non-radial pulsation modes~\citep{wood1}. The most interesting object among pulsating C-rich stars is R Lep, which demonstrates eclipse-like fadings every 18 years~\citep{rlep, rlep2}.

\begin{figure}[h]
\centering
\includegraphics[width=0.9\textwidth]{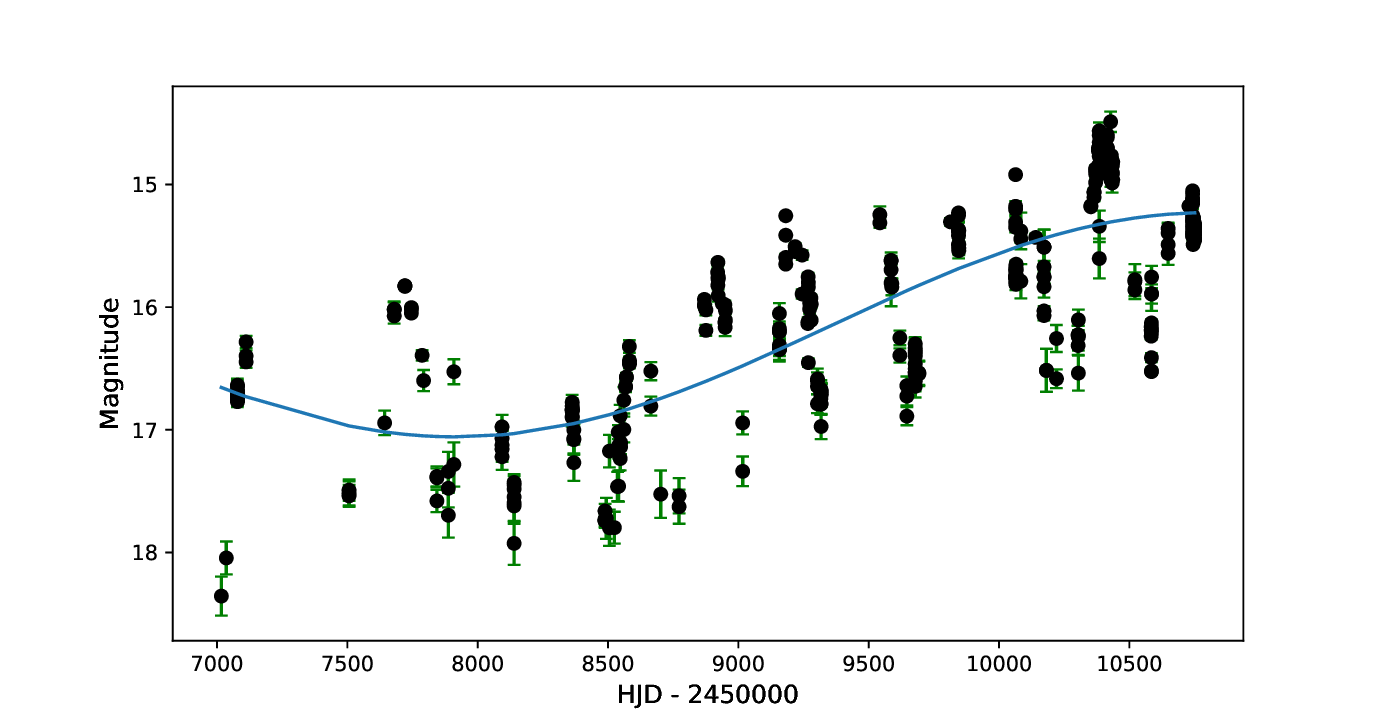}
\caption{Master Net archive light curve of J0701 (black dots) in unfilered band with sinusoidal fit of long term trend with period of 5785 days (blue line).}\label{trend}
\end{figure}

We estimated possible long secondary period of J0701 by fitting the light curve with a sinusoidal model of the form $a+bsin(cT+d)$, where $T$ is time in years. We used a combined light curve from all MASTER network telescopes to achieve better data completeness. The estimated value of long secondary period is $5785\pm428$ days. The light curve with a sinusoidal trend superimposed on it is shown in the fig~\ref{trend}.

Taking into account the fact that J0701 is carbon-rich, we hypothesize that the trend in the light curve may be related to dust formation. This is supported by the presence of an infrared excess with a peak at approximately $2~\mu m$ in the SED J0701, constructed using WISE~\citep{WISE} and 2MASS~\citep{2MASS} data. This is a clear indicator of the presence of a circumstellar dust shell, which is observed in a large number of late-type stars~\citep{tdra,J2101}. An increase in the average brightness level may be due to the dissipation of a shell rich in carbon dust. We did not detect a change in the period, so we do not associate the increase in average brightness with a switching of pulsation modes. Moreover, the Miras are fundamental mode pulsators~\citep{paw}, and this mechanism is not so typical for them. We cannot exclude the presence of a low-mass companion to the star, which causes changes in brightness, however, approximation of a long-term trend with a sinusoid gives the expected period of tens of years.

To finally clarify the nature of variations in the average brightness level, not only photometric observations over decades are required, but also spectroscopic and speckle interferometric ones~\citep{Safonov}. This will allow us to explore the structure of the envelopes around the star and check for possible binarity.

\section{Astrophysical parameters estimation}\label{sect:param}

\begin{figure*}[h]
\centering
\includegraphics[width=0.9\textwidth]{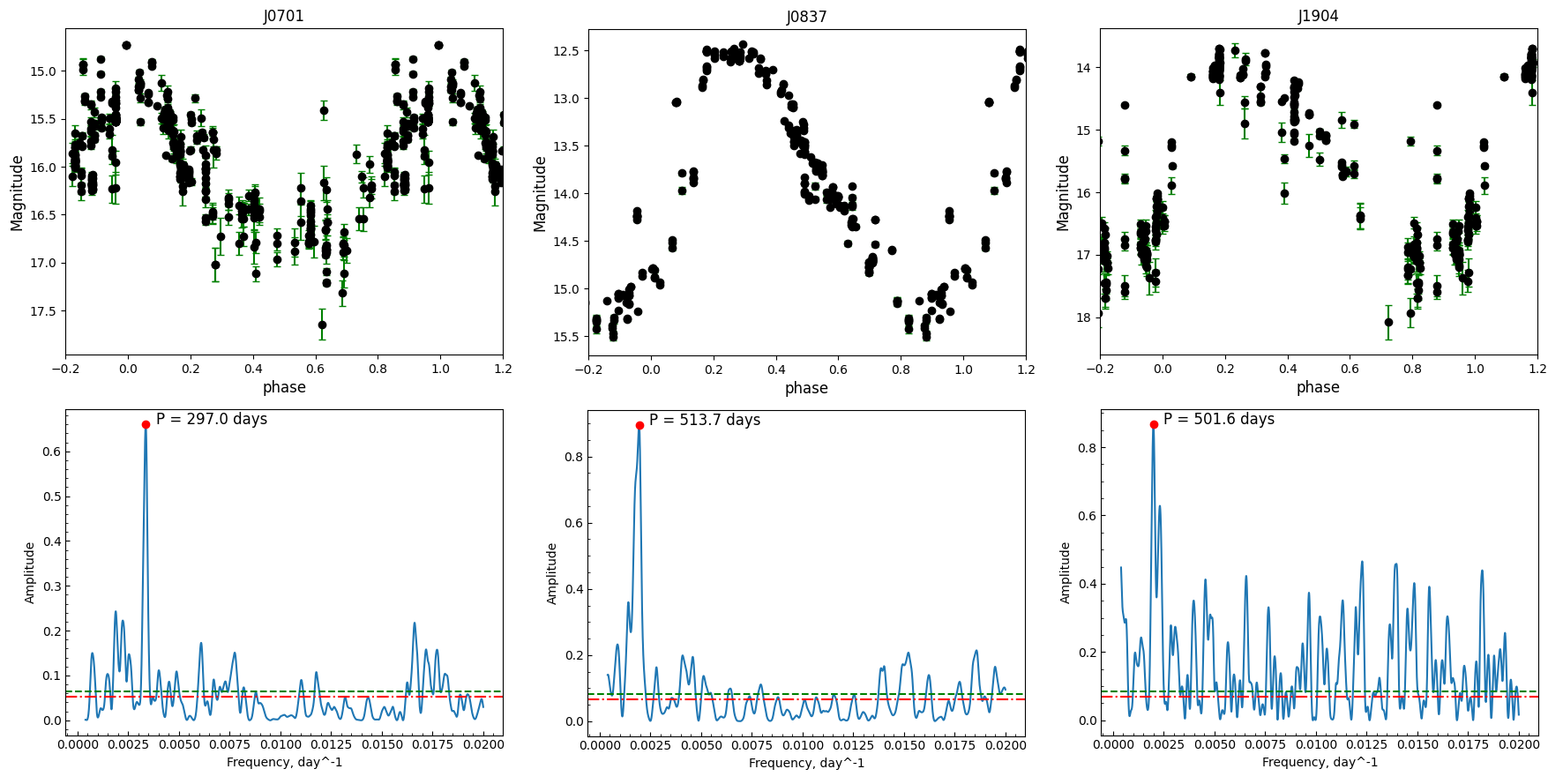}
\caption{Lomb-Scargle periodograms for Mira variables under study (lower panels, green and red dashed lines correspond to false alarm level of 1 and 10 percent respectively) and light curves folded with highest power periods (upper panels). For J0701, detrended data was used.}\label{LSP}
\end{figure*}

To confirm the classification of objects as Mira variables, we used the $UPSILoN$ package~\citep{upsilon}, which gives classification probability of 0.78, 0.76 and 0.71 for J0701, J0837 and J1904. However, this classifier is not adapted to long-period variables, and probabilities may be underestimated. The conclusion about the belonging of stars to Mira variables is made on the basis of comprehensive accounting of all parameters, such as shape of the folded light curves, pulsation periods, infrared $J-K$ color indexes and the correspondences of the relationship between the obtained parameters with the corresponding relationships for Mira variables. To visualize the light curves and check the period estimates obtained with WinEfk, we used an astropy-based tool. Fig.~\ref{LSP} presents Lomb-Scargle periodograms (period interval 50-2500 days) and folded light curves of all three Mira-type variables. It is worth noting that the periods calculated using the Lomb-Scargle periodogram differ from those calculated using WinEfk, but these differences lie within the estimated errors. Since the fast period error estimation method based on the peak FWHM is not an adequate error estimation method~\citep{VdP}, especially in cases of significantly non-Gaussian peaks in our data, we used for evaluation of peak significance false alarm probability and false alarm level calculations and bootstrapping. The false alarm probabilities of the main peaks calculated using $Astropy$ turned out to be many orders of magnitude smaller than 1\%. Therefore, we applied bootstrapping of the data. For each light curve, the periods for 1000 bootstraps were calculated similarly to the original data. Then, the standard deviation of the distribution of the resulting 1000 periods was used to determine the error of period.

For all three discovered Mira-type variables, Gaia parallaxes are determined with large errors, caused by Gaia point spread function chromaticity effect emerging for variable stars with large variations in brightness and temperature~\citep{RcB_gaia}, so the parameters such as luminosity or radius can not be calculated from distance and brightness estimates. Thus, various dependencies of fundamental astrophysical parameters based on statistical analysis of the population of Mira variables and modeling of physical processes in these stars were used. We used period-luminosity relations from~\cite{2023MNRAS.523.2369S} to estimate absolute magnitudes of all three Mira-type variables from periods, obtained from light curves. It is worth noting that most papers on period-luminosity relations for Mira-variables, e. g.~\cite{Feast, 2024RAA....24g5003C} provide these relations for LMC and SMC Mira variables, which may cause systematic errors when applied to galactic Mira variables due to differences in metallicities. ~\cite{2023MNRAS.523.2369S} provides actual period-luminosity relations for both C-rich and O-rich galactic Mira variables based on Gaia DR3 data, therefore its use is optimal in our case. Herewith, we took into account the fact that object J0701 is carbon-rich. The period-absolute magnitude relationship is
\begin{equation}\label{eq1}
m_{\mathrm{abs}}(P) = a_m + 
\begin{cases} 
b_m (\log_{10} P - 2.3), & \text{if } \log_{10} P \leq 2.6, \\
0.3 b_m + c_m (\log_{10} P - 2.6), & \text{else}. 
\end{cases}
\end{equation}
where $m_{abs}$ is the absolute magnitude, $P$ is the period and $a_m$, $b_m$ and $c_m$ are relation coefficients. This dependence provides estimates of absolute stellar magnitude in the generally accepted JHK photometric system, which allows calculating the luminosity using bolometric corrections. We used bolometric corrections from~\cite{BolCor} to convert absolute magnitudes to luminosities. We obtained luminosity estimates of $log_{10}\frac{L}{L_{\odot}}$ $4.11\pm0.13$, $4.14\pm0.13$ and $3.55\pm0.12$ for J0837, J1904 and J0701 respectively.

To estimate the radii of the stars and their masses on ZAMS we used the theoretical dependencies developed by~\cite{Fad23}. These dependencies are
\begin{equation}\label{eq2}
\log \frac{\bar R}{R_{\odot}} = a_0 + a_1 \log \Pi_k, 
\log \frac{\bar L}{L_{\odot}} = b_0 + b_1 \log \Pi_k,
\end{equation}
where $a_0$, $a_1$, $b_0$ and $b_1$ are coefficients, being functions of $M_{ZAMS}$.
From the pulsation periods we obtained the $log_{10}\frac{R}{R_{\odot}}$ values of $2.68\pm0.04$, $2.69\pm0.05$ and $2.33\pm0.04$ and $\frac{M_{ZAMS}}{M_{\odot}}$ estimates $3.0\pm0.1$, $3.1\pm0.1$ and $1.2\pm0.1$ for J0837, J1904 and J0701 respectively.

We compared the obtained parameters of the Mira variables with the dependencies linking the mass, radius and pulsation period from~\cite{1986ApJ...311..864O}. These relations are
\begin{equation}\label{eq3}
logP = -1.92 - 0.73 logM + 1.86 logR
\end{equation}
for fundamental mode pulsators, and
\begin{equation}\label{eq4}
logP = -1.40 + 1.5 logR - 0.5 logM
\end{equation}
for first overtone pulsators. All three Mira type stars (J0701, J0837 and J1904) fit well into these dependencies for fundamental mode pulsators; the periods obtained by substituting the masses and radii we obtained are consistent with those determined from the light curves. Therefore, we can conclude that our methods for determining fundamental astrophysical parameters are reliable, and all three studied stars are fundamental mode pulsators.

It is worth noting that the methods we used do not require knowledge of the absolute magnitude of the stars being studied. Using unfiltered band for photometry does not affect the period value. Typically measurements in the unfiltered band and photometric bands such as Johnson $V$ or SDSS $g$ have a systematic shift on several tenth of magnitude relative to each other~\citep{J0722}. However, the changes in the shape and amplitude of the light curve are minimal, which allows combining the light curves in different bands by a parallel shift, as was done in~\cite{Aigerim}. For Mira variables with large variability amplitudes, such effects are of minimal importance.

\section{Results and conclusions}\label{sect:res}

We present results of red giant variable stars detection and analysis based on detection notices and archival light curves investigation. These stars were found by MASTER global robotic network during regular survey. We also give the description of algorithms of observations, data reduction, and MASTER images storage at Lomonosov supercomputer. Unique duration and duty cycle of observation series from the archive of MASTER Robotic Net can provide accurate photometric time series for long period variable stars. The amount of data accumulated in the MASTER network archive allows for the precise calculation of periods for such objects, and the obtained period values are in good agreement with independently estimated values from other survey projects. Processing MASTER time series allows to study long-term changes in objects such as Mira variables and estimate fundamental astrophysical parameters of such objects: luminosities, radii and ZAMS masses.

We analyzed three optical transients, detected by MASTER automatic transient detecting system, which are connected with variable stars: Mira type variables MASTER OT $J083717.54 - 573411.1$, MASTER OT $J190436.33 + 192828.7$ and
MASTER OT $J07010810 - 6818548$. Archive light curve of J0701 shows mean magnitude growth but constant pulsation period (see Fig.~\ref{Fig3}). It is a spectroscopically validated carbon star. We analyzed possible mechanisms of such behavior of this star. For two other variable objects, detected by MASTER: J0837 and J1904 (both are Mira type variable stars) we presented light curves, period calculation and discovery images with reference ones. We estimated fundamental astrophysical parameters for all three Mira-type variables: luminosities, radii, and ZAMS masses. We used statistical period-luminosity relationships and theoretical dependencies linking radius, period, and luminosity. Validation of the obtained astrophysical parameters and the period-mass-radius relationships confirms the correctness of the chosen strategy and allows classifying the objects under study as fundamental mode pulsators.

Thus, thanks to the presence of an archive of frames for a period of more than twenty years and a system of automatic detection of variable objects, the MASTER network allows us to study not only fast-changing optical transients, but also to analyze slowly-changing variable stars. The integrated use of both theoretical and observed relationships between fundamental parameters of stars allows us to determine and validate their main characteristics (masses, luminosities, radii, pulsation periods) despite the absence of parallax measurements.

\backmatter

\bmhead{Data Availability Statement}

All observational data are available upon reasonable request to the corresponding author.

\bmhead{Acknowledgements}

MASTER equipment was supported by Lomonosov Moscow State University Development Program. The research is carried out using the equipment of the shared research facilities of HPC computing resources at Lomonosov Moscow State University. This research has made use of the Aladin sky atlas \citep{1999ASPC..172..229B} and the VizieR catalog access tool \citep{vizier2000} both developed at CDS, Strasbourg, France. 
This research has made use of the Astrophysics Data System, funded by NASA under Cooperative Agreement 80NSSC21M00561. This research has made use of the ASAS-SN Sky Patrol photometric data~\citep{asassn, asassn2}. The authors are grateful to Dr. V. P. Goranskij for providing his software WinEfk for period search and to A. M. Tatarnikov, N. A. Maslennikova and N. N. Samus for fruitful discussions. The authors express their sincere gratitude to the Gravity Frontiers Foundation for supporting the team and this research. A.Tarasenkov acknowledges the support of the Foundation for the Development of Theoretical Physics and Mathematics BASIS (project 25-2-1-39-1). This research has made use of the Astropy~\citep{AP1, AP2,AP3} Python library. The study was conducted under the state assignment of Lomonosov MSU. MASTER database research is carried out using the equipment of the shared research facilities of HPC computing resources at Lomonosov MSU. The authors thank the anonymous reviewer for valuable comments that helped significantly improve the paper.

\bmhead{Conflict of interest}

The authors declare no conflict of interest.

\bmhead{Funding}

This work was carried out using current institutional funding and did not receive any additional funding.

\bmhead{Ethics approval}

Not Applicable.

\bibliography{sn-bibliography}

\end{document}